\documentclass{PoS}

\usepackage{graphicx,epsfig,wrapfig}
\usepackage{amsmath,amssymb,bm,cite}
\usepackage[dvipsnames]{xcolor}
\usepackage{ulem}
\usepackage{slashed}
\usepackage{color}

\newcommand{\be}{\begin{equation}}
\newcommand{\ee}{\end{equation}}
\newcommand{\ba}{\begin{eqnarray}}
\newcommand{\ea}{\end{eqnarray}}
\newcommand{\la}{\langle}
\newcommand{\ra}{\rangle}
\newcommand{\di}{\!{\rm d}}

\title{Energy-momentum tensor densities in the bag model}

\ShortTitle{Energy-momentum tensor densities in the bag model}

\author{M. J. Neubelt, A. Sampino, J. Hudson, \speaker{K. Tezgin}, P. Schweitzer\\
             Department of Physics, University of Connecticut,
        Storrs, CT 06269, U.S.A.\\
             E-mail: \email{kemal.tezgin@uconn.edu}}

\abstract{
The form factors of the energy-momentum tensor can be accessed via studies of generalized parton distributions in hard exclusive reactions. In this talk we present recent results on the energy-momentum tensor form factors and densities in the bag model formulated in the large-$N_c$ limit. The simplicity and lucidity of this quark model allow us to investigate many general concepts which have recently attracted interest, including  pressure, shear forces and angular momentum density inside the nucleon. The results from the bag model are theoretically consistent, and comply with all general requirements.}

\FullConference{XXVII International Workshop on Deep-Inelastic Scattering and Related Subjects - DIS2019\\
		8-12 April, 2019\\
		Torino, Italy}

\begin{document}

\section{Introduction}

The nucleon form factors of the symmetric 
energy-momentum tensor (EMT) $\hat T^a_{\mu\nu}$  \cite{Kobzarev:1962wt,Pagels:1966zza} can be defined as follows
\begin{align}
    \la p^\prime,s^\prime| \hat T^a_{\mu\nu}(0) |p,s\rangle
    = \bar u(p^\prime)\biggl[A^a(t)\,\frac{P_\mu P_\nu}{M_N}+&
    J^a(t)\ \frac{i(P_{\mu}\sigma_{\nu\rho}+P_{\nu}\sigma_{\mu\rho})
    \Delta^\rho}{2M_N} \nonumber \\
    + &D^a(t)\,
    \frac{\Delta_\mu\Delta_\nu-g_{\mu\nu}\Delta^2}{4M_N}+
    \bar{c}^a(t)\,M_N\,g^{\mu\nu}\biggr]u(p)\, ,
    \label{Eq:ff-of-EMT}
\end{align}
with the kinematical variables defined as
$P=\frac12(p+p')$, $\Delta=(p'-p)$, $t=\Delta^2$.
The form factors for different types of partons ($a=g,\,u,\,d,\,\dots\,$) are renormalization scale dependent which we do not indicate for brevity, whereas the total form factors are renormalization scale independent. EMT form factors satisfy the constraints $\sum_a A^a(0)=1$ and $\sum_a J^a(0)=1/2$ reflecting the fact that the constituents carry the total mass and spin of the particle. The other fundamental property of matter, the D-term defined by $\sum_a D^a(0)=D$, is not constrained. The last form factor satisfies 
$\sum_a \bar{c}^a(t)=0$ due to EMT conservation, $\partial_\mu\hat{T}^{\mu\nu}=0$ where  $\hat{T}^{\mu\nu}=\sum_a\hat{T}^{\mu\nu}_a$. 

The physical content of EMT form factors in terms of densities can be revealed in the Breit frame where the momentum transfer $\Delta^\mu$ is purely spatial. In the Breit frame one can define the static energy-momentum tensor as (with 
$E=\sqrt{M_N^2+\bm{\Delta}^2/4}$) \cite{Polyakov:2002yz}
\begin{equation}\label{Def:static-EMT}
    T_{\mu\nu}({\mathbf{r}},{\mathbf{s}}) =
    \int\frac{\di^3\bm\Delta}{2E(2\pi)^3}\;\exp(-i{\mathbf{r}}\bm{\Delta})\;
    \la p^\prime,S^\prime|\hat{T}_{\mu\nu}(0)|p,S\ra.
\end{equation}
The components of $T_{\mu\nu}({\mathbf{r}},{\mathbf{s}})$ can be interpreted as follows. The $T_{00}({\bf r},{\bf s})$ component corresponds to the energy density inside the nucleon. The components $\epsilon^{i j k} r_j T_{0k}({\bf r},{\bf s})$ correspond to the distribution of angular momentum inside the nucleon. The components of $T_{ik}({\bf r},{\bf s})$ characterize the spatial distributions of pressure and shear forces experienced by the partons inside the nucleon \cite{Polyakov:2002yz,Polyakov:2018zvc}.

\section{EMT form factors in large-$N_c$ limit}
\label{Sec:FFs}

In the large-$N_c$ limit, as the number of colors increases in the color gauge group $SU(N_c)$, the mass of the nucleon increases linearly, $M_N={\cal O}(N_c)$, while the nucleon size $R={\cal O}(N_c^0)$ remains stable \cite{Witten:1979kh}. Hence in the large-$N_c$ limit the nucleon becomes denser, heavier
and it's motion becomes non-relativistic. For the kinematic variables this implies that
$       P^0={\cal O}(N_c)$,
$       \vec{P}={\cal O}(N_c^0)$,
$       \vec{\Delta}={\cal O}(N_c^0)$,
$        \Delta^0={\cal O}(N_c^{-1})$.

On the other hand, the EMT form factors have the following scaling behavior in the large-$N_c$ limit for the isoscalar $(Q=u+d)$ flavor combinations \cite{Goeke:2001tz} 
\be\label{Eq:EMT-ff-large-Nc}
        A^{Q}(t)  ={\cal O}(N_c^0),       \quad
        J^{Q}(t)  ={\cal O}(N_c^0),       \quad
        D^{Q}(t)  ={\cal O}(N_c^2),       \quad
        \bar{c}^{Q}(t)  ={\cal O}(N_c^0).
\ee
Considering the kinematic variables in the large-$N_c$ limit 
and (\ref{Eq:EMT-ff-large-Nc}), the EMT form factors (\ref{Eq:ff-of-EMT}) in the large-$N_c$ limit become 
\begin{align}
\langle p^\prime,s^\prime| \hat T^{00}_{Q}(0) |p,s\rangle
        = 2\,M_N^2\biggl[
        & A^{Q}(t) - \frac{t}{4M_N^2}\,D^{Q}(t)
        + {\bar c}^{Q}(t)\biggr]\delta_{ss^\prime}
        \label{Eq:EMT-Breit-T00}\\
\langle p^\prime,s^\prime| \hat T^{ik}_{Q}(0) |p,s\rangle
        = 2\,M_N^2\biggl[
        & D^{Q}(t)\,\frac{\Delta^i\Delta^k-\delta^{ik}{\bm{\Delta}}^2}{4M_N^2}
        - {\bar c}^{Q}(t)\,\delta^{ik} \biggr] \delta_{ss^\prime}
        \label{Eq:EMT-Breit-Tik}\\
\langle p^\prime,s^\prime| \hat T^{0k}_{Q}(0) |p,s\rangle
        = 2\,M_N^2\biggl[
        & J^{Q}(t)\;\frac{(-i\,\bm{\Delta}\times\bm{\sigma}_{s^\prime s})^k}{2\,M_N}
        \biggr]
        \label{Eq:EMT-Breit-T0k}
\end{align}
where $\delta_{ss'}$ and $\bm{\sigma}_{s^\prime s}$ denote $\chi^\dag_{s'}\chi^{ }_s$ and $\chi^\dag_{s'}\bm{\sigma}\chi^{ }_s$, respectively, and $\chi_s$ denotes the nucleon two-spinor.

\section{EMT form factors in the bag model} \label{FFsBM}
In this section we present results \cite{Neubelt:new} for the form factors in the large-$N_c$ limit in the bag model \cite{Chodos:1974je}. In this model one describes the nucleon by placing $N_c=3$ non-interacting quarks in a color-singlet state inside a "bag" with boundary conditions to confine quarks inside the bag. In its rest frame the bag is a spherical region of radius $R$ carrying a positive energy density $B$. The bag model EMT form factors have first been studied by Ji et al.~\cite{Ji:1997gm} for finite $N_c$. In our approach we
choose to work in the large-$N_c$ limit where the motion of the heavy nucleon becomes non-relativistic. 

\begin{figure}[b!]
\centering
\includegraphics[width=3.8cm]{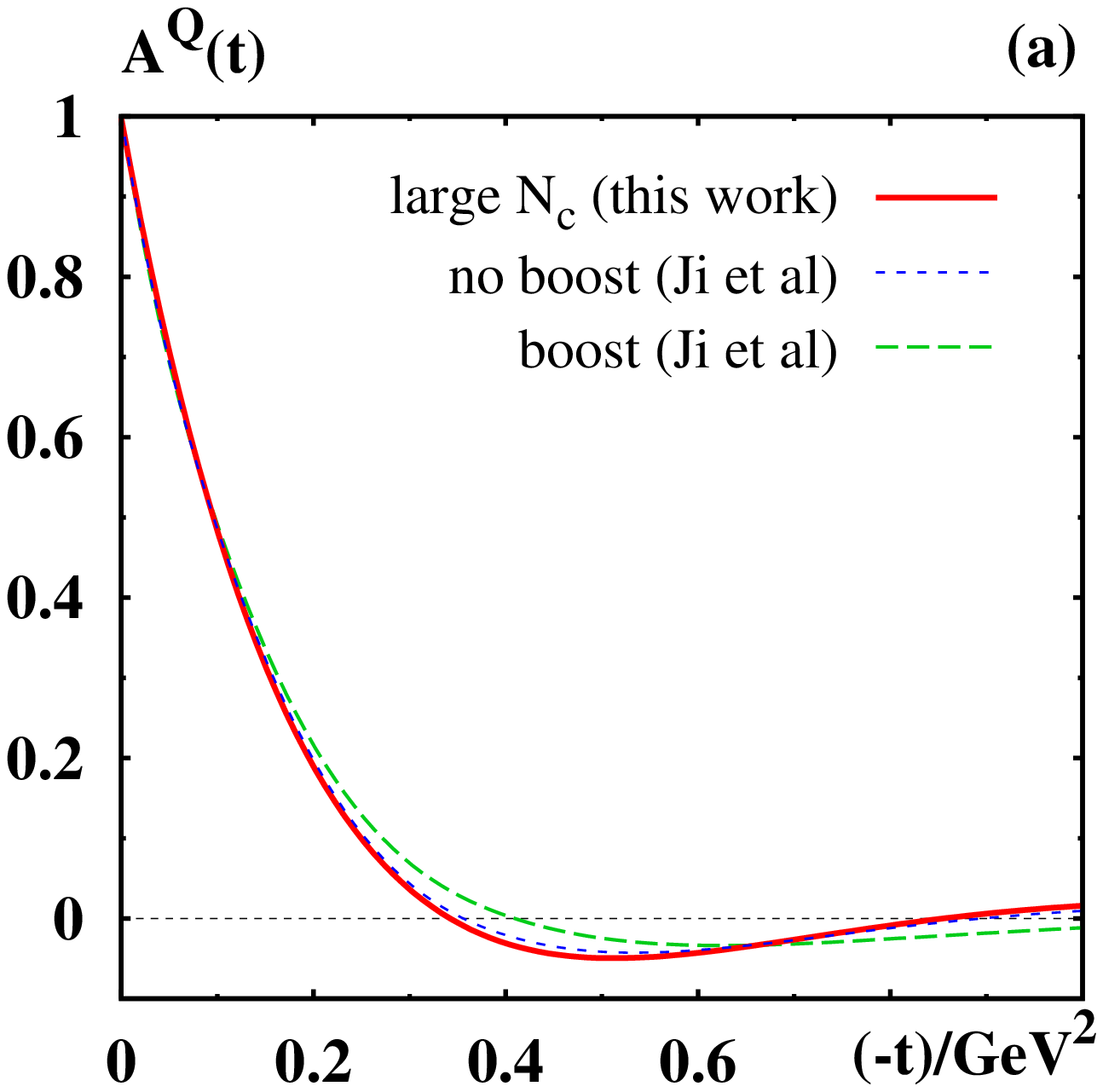}%
\includegraphics[width=3.8cm]{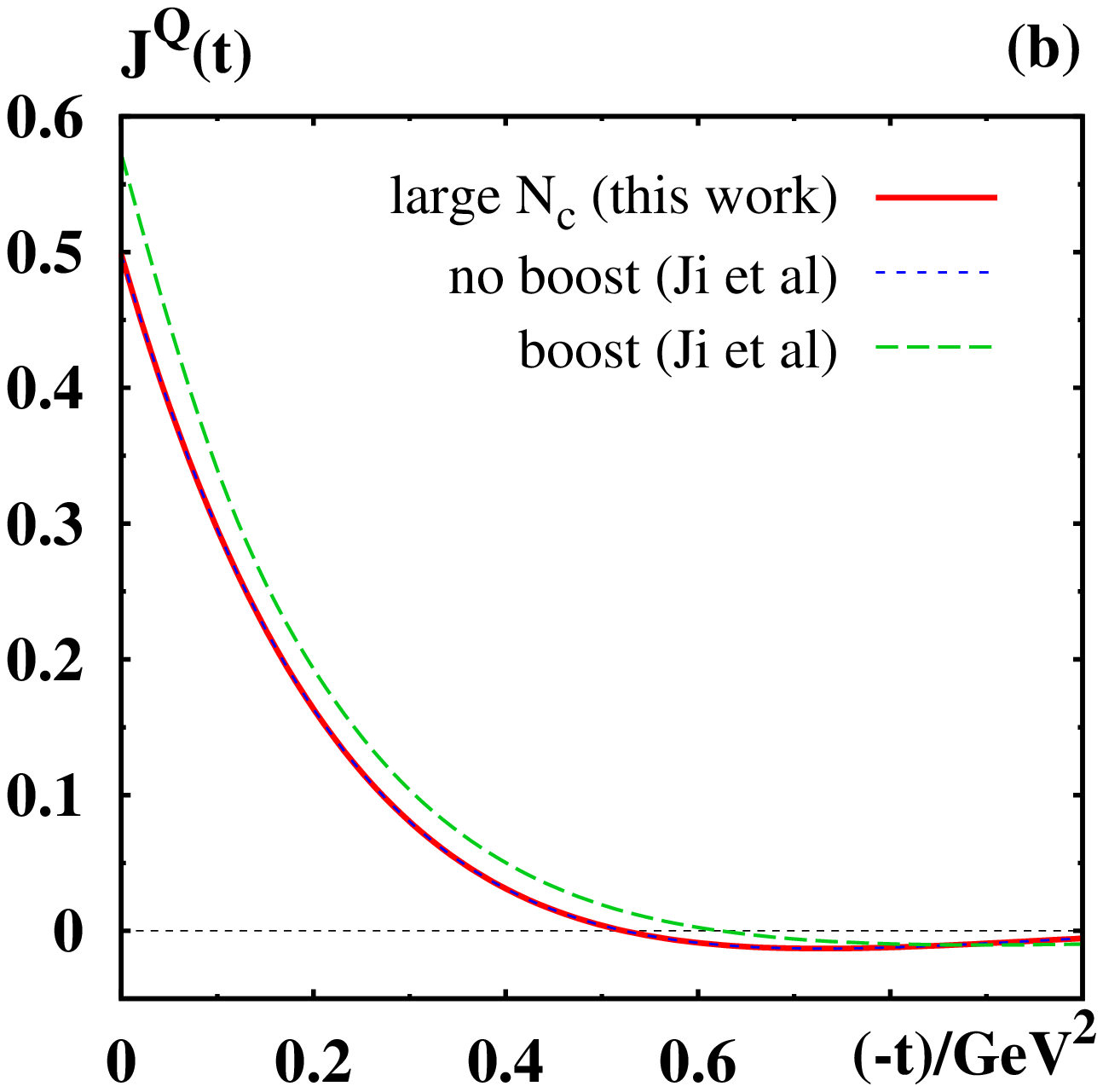}%
\includegraphics[width=3.8cm]{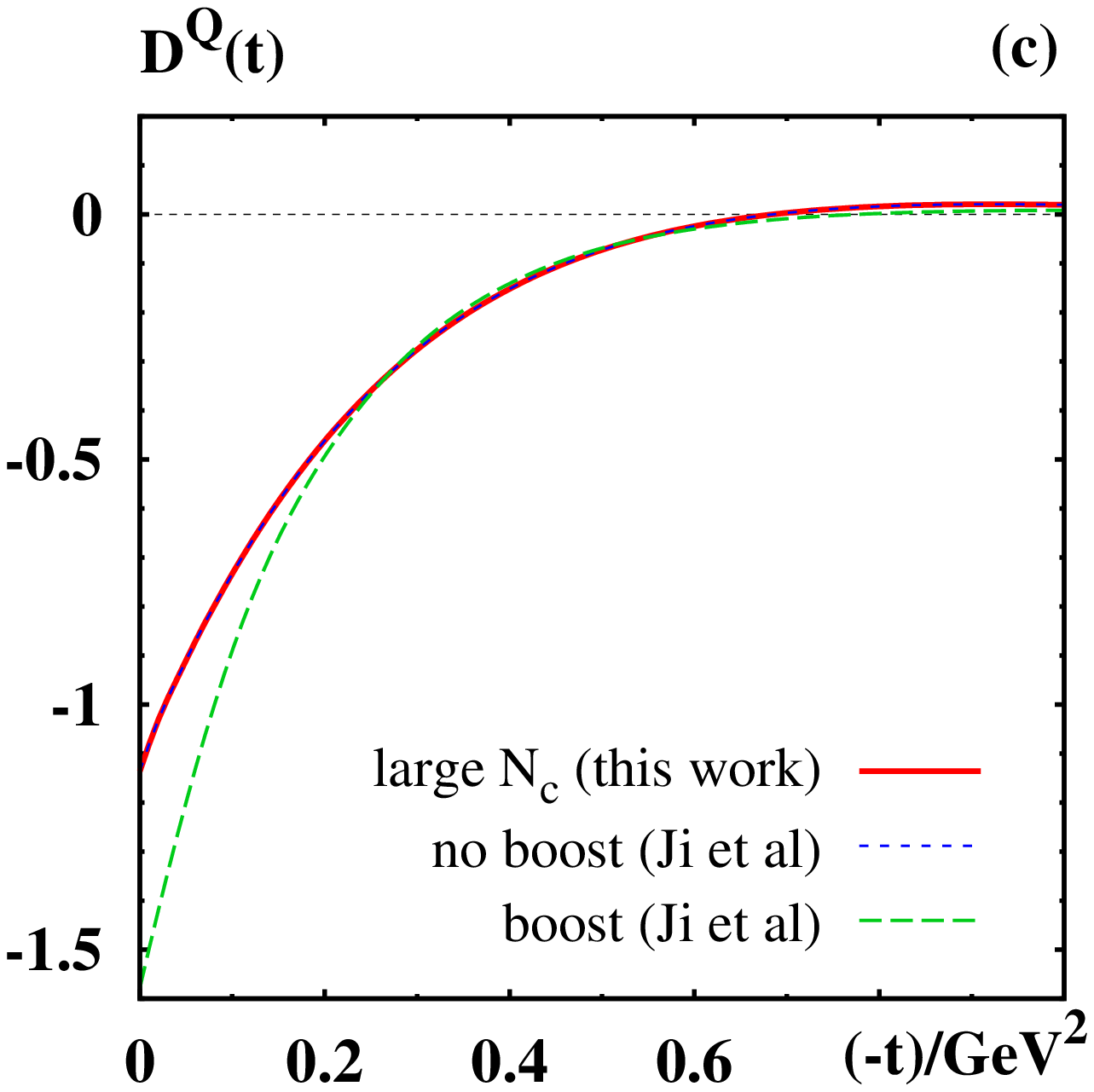}%
\includegraphics[width=3.8cm]{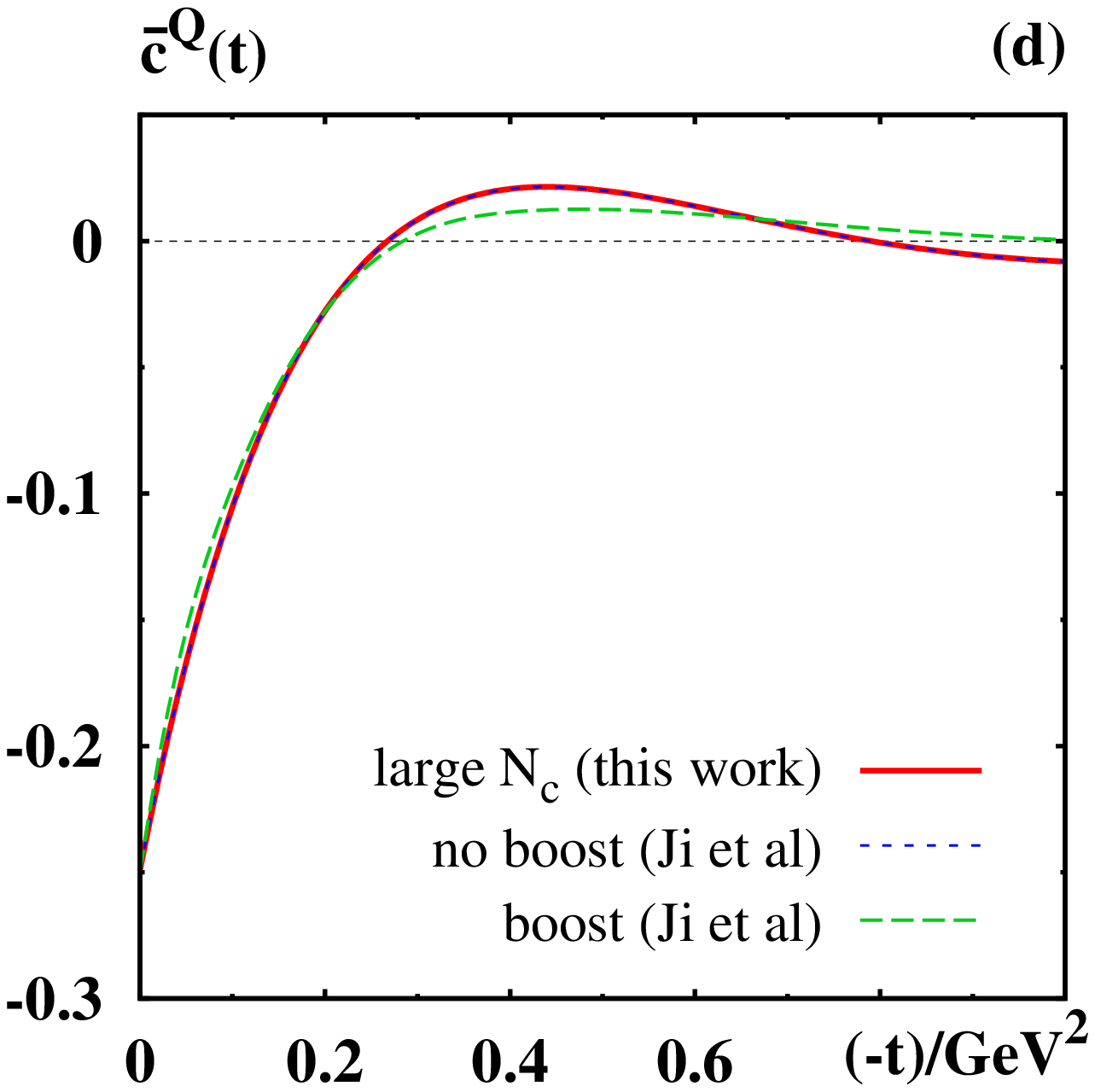}
\caption{\label{Fig-01:FFs} EMT form factors in the bag model for the contribution of quarks ($Q=u+d$) in the large $N_c$ limit (solid lines, this work). For comparison we also show results by Ji et al., Ref.~\cite{Ji:1997gm}, computed in the bag model without (dotted lines) and with (dashed lines) considering boosts \cite{Neubelt:new}.}
\end{figure}

The EMT in bag model has quark and bag contributions. 
The symmetric quark EMT operator is given by
\be\label{Eq:symmetricEMT}
        T^{\mu\nu}_q = \frac{1}{4}\overline{\psi}_q\biggl(
        -i\overset{ \leftarrow}\partial{ }^\mu\gamma^\nu
        -i\overset{ \leftarrow}\partial{ }^\nu\gamma^\mu
        +i\overset{\rightarrow}\partial{ }^\mu\gamma^\nu
        +i\overset{\rightarrow}\partial{ }^\nu\gamma^\mu\biggr)\psi_q\,,
\ee
where the arrows indicate which wave functions are differentiated.
Evaluating the operator (\ref{Eq:symmetricEMT}) in the bag model yields the form factors depicted in Figure \ref{Fig-01:FFs}. As the figure shows, the whole momentum and spin of the nucleon is carried by the quarks since $A^{Q}(0)=1$ and $J^{Q}(0)=1/2$. The D-term is, on the other hand, not constrained and we obtain the value $D^{Q}=-1.145$ for massless quarks. The last form factor $\bar{c}^{Q}(t)$ is non zero due to the fact that the quark EMT (\ref{Eq:symmetricEMT}) is not conserved. Taking into account the bag contribution one finds that the constraint 
$\sum_a \bar{c}^a(0)=0$ is satisfied. The results shown in Figure \ref{Fig-01:FFs} refer to the large-$N_c$ limit and are consequently valid for $|t|\ll M_N^2$. Considering the large-$N_c$ limit the
results agree with \cite{Ji:1997gm}.

\section{EMT densities in the bag model}
In order to compute EMT densities we need to take the Fourier transform of $\la p^\prime,S^\prime|\hat{T}_{\mu\nu}(0)|p,S\ra$ with respect to $\bm\Delta$ as expressed in (\ref{Def:static-EMT}). In bag model one can also directly evaluate the matrix elements of the EMT in coordinate space. Both approaches yield the same result for quarks \cite{Neubelt:new}. 

\begin{figure}[b!]
\begin{centering}
\includegraphics[width=4.8cm]{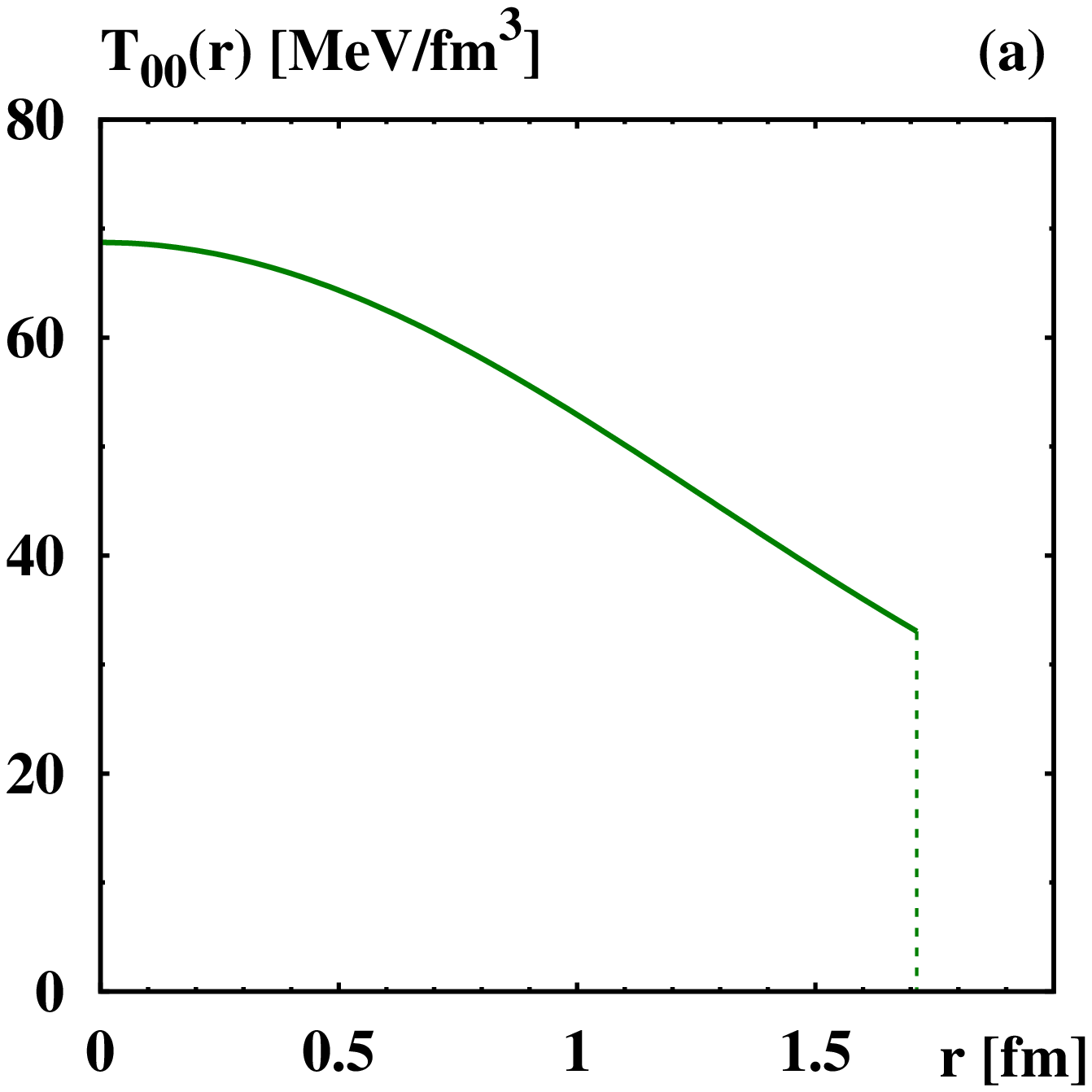} \
\includegraphics[width=4.8cm]{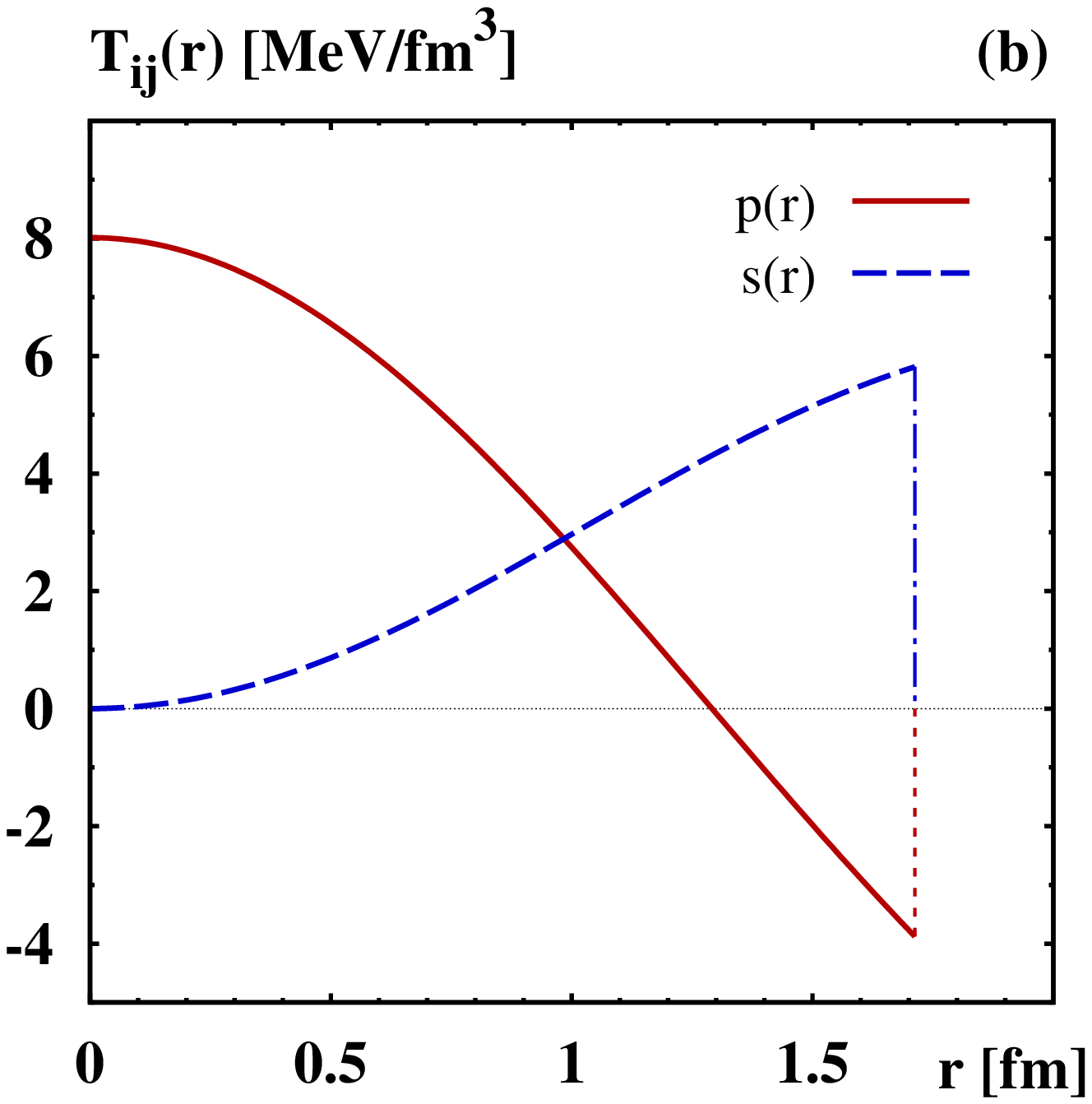} \
\includegraphics[width=4.8cm]{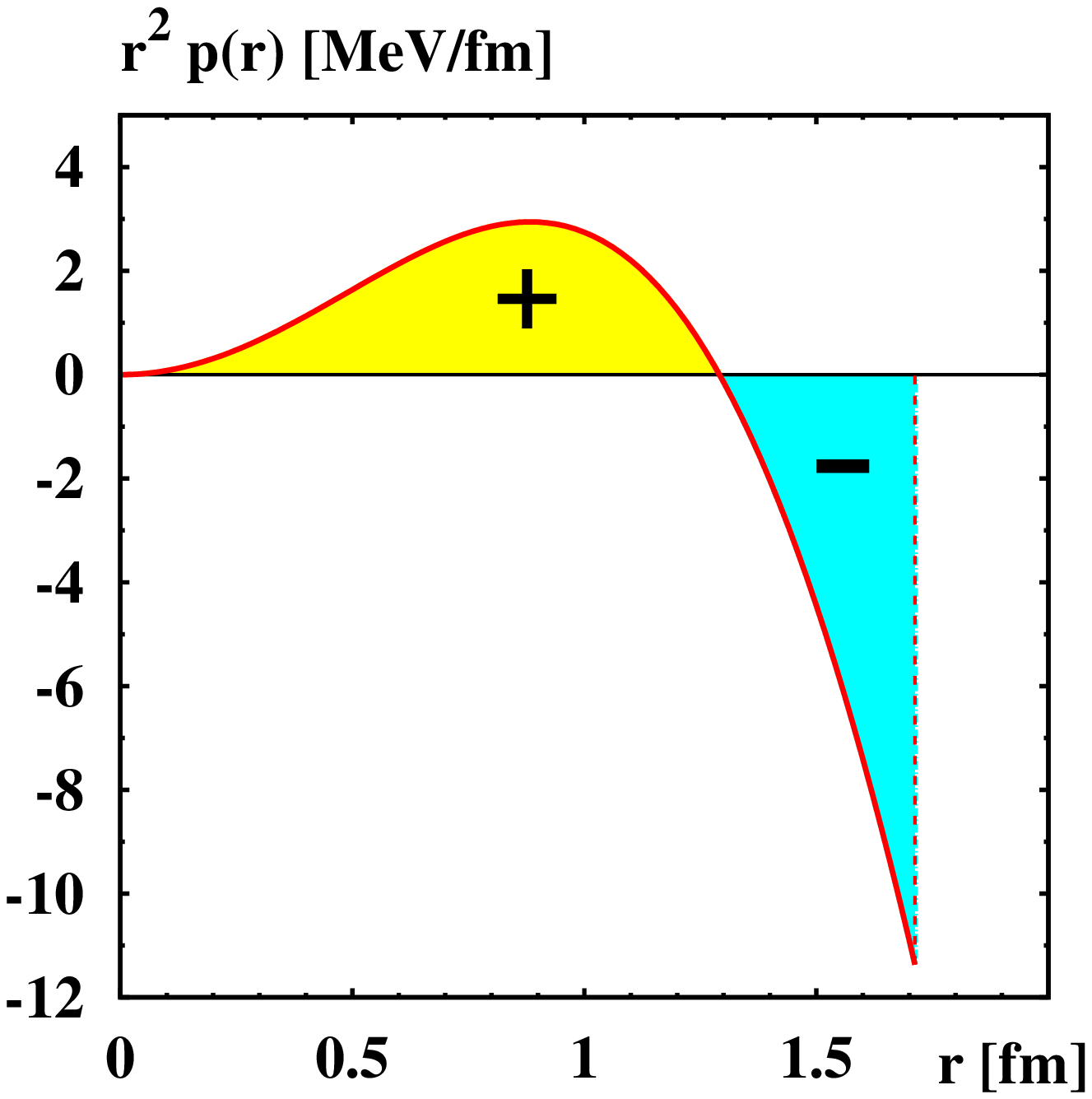}
\par\end{centering}
\caption{\label{Fig-2:T00-p-s} 
The total ($Q+\,$bag) contributions to
(a) energy density $T_{00}(r)$ and
(b) stress tensor $T_{ij}(r)$ densities $s(r)$ and $p(r)$ as functions of $r$ in the bag model. 
(c) Illustration how the von Laue condition is realized in 
the bag model: $r^{2}p(r)$ as function of $r$. 
The areas above and below the $r$-axis are equal and compensate each
other in the integral $\int_{0}^{R}dr\;r^{2}p(r)=0$ according to
Eq.~(\ref{Eq:stability}).
The results refer to massless quarks.
The vertical lines mark the position of the bag boundary 
at $R\approx1.71\,$fm.}
\end{figure}

The energy density $T_{00}(\bf r)$ receives contributions both from quarks and the bag density $B$. The quarks (the bag) contribute $\frac{3}{4}$ ($\frac{1}{4}$) of the total nucleon mass for any $N_c$
in massless case. We show our numerical result for $T_{00}(\bf r)$ in Figure \ref{Fig-2:T00-p-s}(a).  
The components $T_{0k}(\bf r)$ receive a contribution 
only from quarks. The angular momentum density is defined by $J_q^i(\mathbf{r})=\epsilon^{ijk} r^j T^{0k}_q(\mathbf{r})$ and receives monopole and quadrupole contributions \cite{Lorce:2017wkb}
which are related to each other in a model independent way \cite{Schweitzer:2019kkd} which holds also in the bag model.
The
$T_{ij}(\mathbf{r})$ components of the static EMT encode the information on the distribution of pressure and shear forces and can be decomposed as
\be\label{Eq:T_ij-pressure-and-shear}
    T_{ij}({\bf r})
    = s(r)\left(\frac{r_ir_j}{r^2}-\frac 13\,\delta_{ij}\right)
        + p(r)\,\delta_{ij}\, , \ee
where the trace part $p(r)$ describes the radial distribution of the ``pressure'', and the traceless part $s(r)$ describes the distribution of the ``shear forces'' inside the nucleon \cite{Polyakov:2002yz}.  Both functions are related to each other due to the EMT conservation by the differential equation
\be\label{Eq:p(r)+s(r)}
    \frac23\;\frac{\partial s(r)}{\partial r\;}+
    \frac{2s(r)}{r} + \frac{\partial p(r)}{\partial r\;} = 0\;.
\ee
Another important property which can be directly derived from the conservation of the EMT is the von Laue relation
\be\label{Eq:stability}
    \int\limits_0^\infty \!\di r\;r^2p(r)=0 \;.
\ee
Notice that the von Laue relation requires the pressure distribution inside the nucleon to change sign, at least once. In bag model $p(r)$ receives  contributions  both  from  quarks  and  the  bag  density B, whereas $s(r)$ receives contribution  only  from  quarks. We depict our numerical results for $p(r)$ and $s(r)$ in Figure \ref{Fig-2:T00-p-s}(b) and for the von Laue relation in Figure \ref{Fig-2:T00-p-s}(c). Notice that in general $p(r)$, $s(r)$ as well as other EMT densities do not vanish at the bag boundary. The bag model results satisfy the differential equation (\ref{Eq:p(r)+s(r)}) as well as the von Laue relation (\ref{Eq:stability}). First phenomenological insights on the pressure distribution inside the nucleon were discussed in \cite{Burkert:2018bqq, Kumericki:2019ddg}.

The conservation of the EMT also provides two equivalent expressions for the $D$-term,
\ba\label{Eq:d1-from-s(r)-and-p(r)}
        D &=& -\,\frac{4}{15}\;M_N \int\di^3{\bf r}\;r^2\, s(r)
         =     M_N \int\di^3{\bf r}\;r^2\, p(r)\;,
\ea
which both yield the result $D=-1.145$ for massless quarks and $N_c=3$, in agreement with numerical results for the form factor $D(t)$
at $t=0$, cf.\ Sec.~\ref{Sec:FFs} and \cite{Neubelt:new, Ji:1997gm}.

\section{Conclusions}

We have studied EMT form factors and their densities in the large-$N_c$ 
limit in the bag model. We have shown that the bag model results 
comply with all general properties of the EMT form factors and densities, 
and exhibit similar features as observed in earlier theoretical studies 
\cite{Goeke:2007fp,Goeke:2007fq,Wakamatsu:2007uc,Cebulla:2007ei,Mai:2012yc,Mai:2012cx,Cantara:2015sna,Hagler:2003jd,Gockeler:2003jfa,Hagler:2007xi,Pasquini:2014vua,Kim:2012ts,Jung:2014jja,Shanahan:2018pib}.
Thus, the bag model is an attractive and simple theoretical playground 
which provides an internally consistent description of the nucleon, 
and allows to investigate in detail the generic EMT of hadrons
\cite{Neubelt:new}. This encourages to explore this model for further
studies of the newly introduced EMT concepts \cite{Polyakov:2018exb,Lorce:2018egm,Polyakov:2018rew,Polyakov:2019lbq,Cosyn:2019aio,Cotogno:2019xcl}. An interesting application of the bag model was
reported in \cite{Hudson:2017oul} where it was shown in general 
that the $D$-term of a fermion vanishes in the free-field case.
The bag model was used in \cite{Hudson:2017oul} to provide an
internally consistent dynamical example illustrating
how the $D$-term of a fermionic system vanishes when
interactions are absent.

\ \\
{\bf Acknowledgments.}
This work was supported by NSF grant no. 1812423 and DOE grant
no. DE-FG02-04ER41309.

\end{document}